\documentclass[aip, jap, reprint, amsmath, amssymb]{revtex4-1}

\usepackage{graphicx}
\usepackage{dcolumn}
\usepackage{bm}
\usepackage{changes}
\usepackage{here}

\begin{document}


\title[Ion distribution and ablation depth measurements of a fs-ps laser-irradiated solid tin target]{Ion distribution and ablation depth measurements of a fs-ps laser-irradiated solid tin target}

\author{M.J.Deuzeman}
 \email{m.j.deuzeman@arcnl.nl}
\affiliation{Advanced Research Center for Nanolithography (ARCNL), Science Park 110, 1098 XG Amsterdam, The Netherlands}
\affiliation{Zernike Institute for Advanced Materials, University of Groningen, Nijenborgh 4, 9747 AG Groningen, The Netherlands}

\author{A.S.Stodolna}
\affiliation{Advanced Research Center for Nanolithography (ARCNL), Science Park 110, 1098 XG Amsterdam, The Netherlands}

\author{E.E.B.Leerssen}
\affiliation{Advanced Research Center for Nanolithography (ARCNL), Science Park 110, 1098 XG Amsterdam, The Netherlands}

\author{A.Antoncecchi}
\affiliation{Advanced Research Center for Nanolithography (ARCNL), Science Park 110, 1098 XG Amsterdam, The Netherlands}

\author{N.Spook}
\affiliation{Advanced Research Center for Nanolithography (ARCNL), Science Park 110, 1098 XG Amsterdam, The Netherlands}
\affiliation{Van der Waals-Zeeman Instituut, University of Amsterdam, Science Park 904, 1098 XH Amsterdam, The Netherlands}

\author{T.Kleijntjens}
\affiliation{Advanced Research Center for Nanolithography (ARCNL), Science Park 110, 1098 XG Amsterdam, The Netherlands}

\author{J.Versluis}
\affiliation{FOM Institute AMOLF, Science Park 104, 1098 XG Amsterdam, The Netherlands}

\author{S.Witte}
\affiliation{Advanced Research Center for Nanolithography (ARCNL), Science Park 110, 1098 XG Amsterdam, The Netherlands}
\affiliation{Department of Physics and Astronomy, Vrije Universiteit, De Boelelaan 1081, 1081 HV Amsterdam, The Netherlands}

\author{K.S.E.Eikema}
\affiliation{Advanced Research Center for Nanolithography (ARCNL), Science Park 110, 1098 XG Amsterdam, The Netherlands}
\affiliation{Department of Physics and Astronomy, Vrije Universiteit, De Boelelaan 1081, 1081 HV Amsterdam, The Netherlands}

\author{W.Ubachs}
\affiliation{Advanced Research Center for Nanolithography (ARCNL), Science Park 110, 1098 XG Amsterdam, The Netherlands}
\affiliation{Department of Physics and Astronomy, Vrije Universiteit, De Boelelaan 1081, 1081 HV Amsterdam, The Netherlands}

\author{R.Hoekstra}
\affiliation{Advanced Research Center for Nanolithography (ARCNL), Science Park 110, 1098 XG Amsterdam, The Netherlands}
\affiliation{Zernike Institute for Advanced Materials, University of Groningen, Nijenborgh 4, 9747 AG Groningen, The Netherlands}

\author{O.O.Versolato}
\affiliation{Advanced Research Center for Nanolithography (ARCNL), Science Park 110, 1098 XG Amsterdam, The Netherlands}

\date{\today}

\begin{abstract}
The ablation of solid tin surfaces by an 800-nanometer-wavelength laser is studied for a pulse length range from 500 fs to 4.5 ps and a fluence range spanning 0.9 to 22 J/cm\textsuperscript{2}. The ablation depth and volume are obtained employing a high-numerical-aperture optical microscope, while the ion yield and energy distributions are obtained from a set of Faraday cups set up under various angles. We found a slight increase of the ion yield for an increasing pulse length, while the ablation depth is slightly decreasing. The ablation volume remained constant as a function of pulse length. The ablation depth follows a two-region logarithmic dependence on the fluence, in agreement with the available literature and theory. In the examined fluence range, the ion yield angular distribution is sharply peaked along the target normal at low fluences but rapidly broadens with increasing fluence. The total ionization fraction increases monotonically with fluence to a 5-6\% maximum, which is substantially lower than the typical ionization fractions obtained with nanosecond-pulse ablation. The angular distribution of the ions does not depend on the laser pulse length within the measurement uncertainty. These results are of particular interest for the possible utilization of fs-ps laser systems in plasma sources of extreme ultraviolet light for nanolithography.
\end{abstract}

\maketitle

\section{\label{sec:intro}Introduction}

Ultrafast lasers, with pulse durations in the femtosecond-picosecond range, are used in a wide range of applications, such as micromachining, thin film deposition, material processing, surface modification, and ion beam generation \cite{kruger1999,anisimov1993, chichkov1996, pronko1995, preuss1995, obona2011, kato2008, doring2010, varel1997}. More recently, these lasers have attracted attention for their possible applicability in the field on tin-based plasma sources of extreme ultraviolet (EUV) light for nanolithography. There they could be used for generating a fine-dispersed liquid-metal target\cite{vinokhodov2016} before the arrival of a high-energy main-pulse responsible for the EUV emission, enhancing laser-plasma coupling\cite{sullivan2015}. The utilization of a fs-ps laser system could strongly reduce fast ionic and neutral debris from EUV sources compared with nanosecond-pulses\cite{toftmann2013}, enabling better machine lifetime\cite{banine2011}. 

Since the 1990s many experiments have been performed and models developed\cite{chichkov1996,anisimov1993} for laser-matter interaction at this particular time scale. Target materials used are metals such as gold, silver, copper and aluminum \cite{amoruso2000, amoruso2002, toftmann2013,nolte1997,verhoff2012,furusawa1999, preuss1995, zhang2002, albert2003, yang2006}, and non-metals such as silicon \cite{shaheen2014, hebeisen2008, menendez2010, amoruso2005a} and metal oxides \cite{mero2005, stoian2000, stoian2002}, among others \cite{ditmere1997, claeyssens2002}. Most of these studies are conducted in a femtosecond pulse length range from 50 fs up to approximately 1 ps and a pulse fluence up to 10 J/cm\textsuperscript{2}. In almost all studies the wavelength of the laser is in the infrared, where commercial laser systems are readily available. The focus is often either on ablation depth or ion distributions (energy, yield or angular), with a few exceptions such as the work of Toftmann et al.\cite{toftmann2013} which addresses both. A detailed study of laser ablation of the relevant element tin, including both depth and ion emission distribution, has not yet been performed in the fs-ps domain. Such a study, however, is indispensable for exploring EUV plasma sources in the short-pulse regime.

In this work, we present a systematic study of the laser ablation of a solid tin target by an 800-nanometer-wavelength laser. We determine the angle-resolved yield and energy distributions of the produced plasma ions through time-of-flight techniques. The depth of the ablation crater was established in addition to the ion measurements using a high-numerical-aperture optical microscope. We varied the laser pulse length between 500 fs and 4.5 ps. In this pulse length range lies a transition regime in which the transfer of laser energy from the heated electrons to the lattice starts playing a significant role\cite{pronko1995, chichkov1996}. Recent work using ultrafast laser pulses to irradiate molten-tin microdroplets hinted at a dramatic change in laser-metal coupling at 800 fs pulse length, resulting in a simultaneous sharp increase in droplet expansion velocity\cite{vinokhodov2016} and a strong dip in the yield of fast ionic debris\cite{vinokhodov2016b}. This makes it highly desirable to provide further data in this pulse length regime. In our experiments, we additionally study the influence of pulse fluence in detail, covering a range from 0.9 to 22 J/cm\textsuperscript{2}, similar as in refs.\,\onlinecite{vinokhodov2016,vinokhodov2016b}. At the high end of this fluence range, the total volume of ablated material reaches $\sim$10\textsuperscript{4} $\mu$m\textsuperscript{3}, which is similar to the volume of a tin droplet used in state-of-the-art plasma sources of EUV light and therefore provides an interesting comparison.

\begin{figure}[b!]
\includegraphics[width=8.5cm]{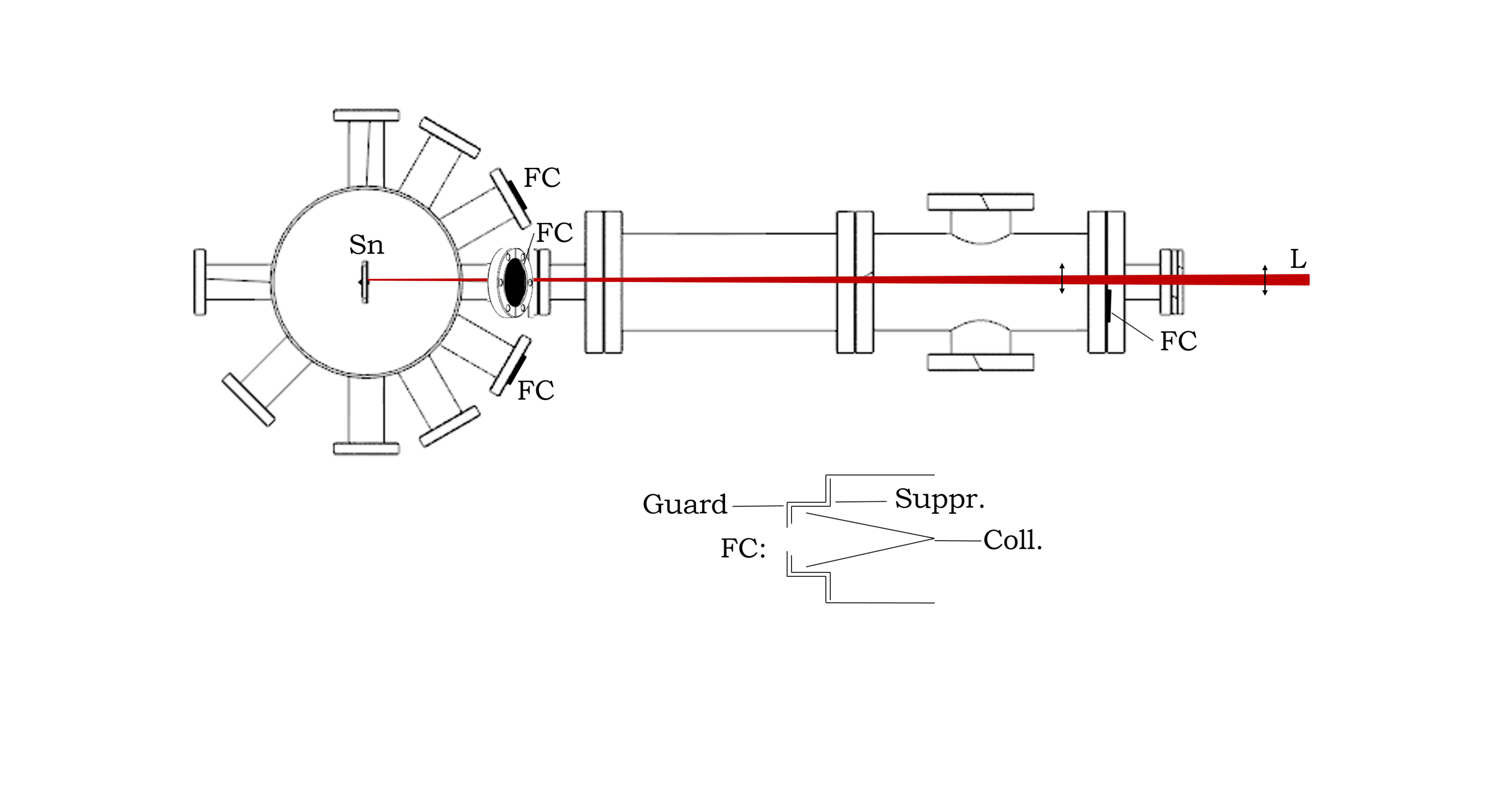}
\caption{The outline of the setup (top view) used for the experiments. The four dark black spots mark the positions of the Faraday cups (FCs): one at 2\textsuperscript{$\circ$} and three at 30\textsuperscript{$\circ$} with respect to the normal of the target. Two of the 30\textsuperscript{$\circ$}-FCs and the 2\textsuperscript{$\circ$}-FC are in the horizontal plane, one of the 30\textsuperscript{$\circ$}-FCs is out of plane. The laser beam (red), horizontally polarized, is incident on the target under normal angle. A schematic cut-through of a home-made FC is also shown. The outer guard shield has a diameter of 6 mm, the inner suppressor shield a diameter of 8 mm. The ion currents are obtained from the collector cone. }
\label{fig:expsetup}
\end{figure}

\section{\label{sec:expset}Experimental Setup}

A solid planar polycrystalline 99.999\% pure tin target with a 1-millimeter-thickness is irradiated by a pulsed 800-nanometer-wavelength Ti:Sapphire laser (Coherent Legend USP HE). The laser beam is incident on the target at normal incidence. The target and detectors is kept at a vacuum of 10\textsuperscript{-8} mbar. The laser pulses have a Gaussian-shaped temporal and spatial profile. All pulse lengths presented in this work are the full-width at half-maximum (FWHM) of the pulse in time-domain. The pulse duration has been changed between 500 fs and 4.5 ps by varying the group velocity dispersion in the compressor of the amplified laser system. The resulting pulse duration was measured using a single-shot autocorrelator. The beam profile of the pulses are slightly elliptical, with a FWHM of 105$\pm$5 $\mu$m on the long axis and 95$\pm$5 $\mu$m on the short axis. The peak fluence, the maximum fluence attained in the center of the Gaussian pulse, is calculated using these widths and the pulse energy. This fluence is varied with a $\lambda$/2 wave plate in combination with a thin-film polarizer, which leaves the spatial profile of the laser beam unchanged. The pulse repetition rate of 1 kHz is reduced with pulse-picking optics to an effective rate of 5 Hz to enable shot-to-shot data acquisition and controlled target movement between the laser pulses. The polarization of the laser light is horizontal (see Fig. \ref{fig:expsetup}). As the pulses are incident on the target at normal incidence, no dependence on the polarization is expected.

Time-of-flight (TOF) ion currents are obtained from Faraday cups (FCs) set up around the target, one at 2\textsuperscript{$\circ$} from the surface normal and at a distance of 73 cm, two at 30\textsuperscript{$\circ$} and 26 cm (in horizontal and vertical position) and one at 30\textsuperscript{$\circ$} and 24 cm (also in the horizontal plane). Three FCs are home-made and consist of a grounded outer guard shield, an inner suppressor shield, and a charge-collector cone (cf. inset in Fig. \ref{fig:expsetup}). A voltage of -100 V on the suppressor shield inhibits stray electrons entering the collector cone and secondary electrons, which may be produced by energetic or multi-charged ions \cite{bodewits2014}, from leaving it. To further reduce the chance of stray electrons arriving at the collector, a bias voltage of -30 V is applied to the collector cone itself. The other FC (at 30\textsuperscript{$\circ$} and 24 cm) has a different design (model FC-73A from Kimball Physics) and can be used for retarding field analysis. Checks with retarding grids using this FC indicate that ions with energies below 100 eV, the vast majority of the ions, are mostly singly charged. The charge yield measured with a FC can thus be regarded as a direct measure of the ion yield. In the conversion from a TOF- to a charge versus the ion energy-signal, the signal is corrected for the non-constant relation between bin size in the time- and in the energy-domain using

\begin{equation}
S_{E}=|\frac{dt}{dE}|S_{t}=\frac{t}{2E}S_{t},
\end{equation}

\noindent in which $S_E$ and $S_t$ are the signals in respectively the energy domain and the time domain, and \textit{t} and \textit{E} respectively denote the TOF and the ion energy. Signals are corrected for the solid angle of the detectors and for the finite response $RC$-time of the circuit. The total charge yields are determined by integrating the charge over the full spectrum. Unless otherwise specified, we use the average of the total charge yield for the three 30\textsuperscript{$\circ$}-FCs. 

To enable depth measurements and to prevent severe target modification by the laser, which would influence the measurements, the target is moved after every 30 pulses. The first pulses on a fresh spot on the target generate signals with a small TOF, indicative of light elements or high-energy tin atoms. Early studies, employing ion energy analyzers, identify these pulses as light elements contaminations\cite{toftmann2013,amoruso2002}. Energy-dispersive X-ray spectroscopy measurements reveal that areas on our tin target unexposed to laser light contain a substantial amount of oxygen and other low-mass elements, such as carbon and nitrogen. These elements are only barely visible, if at all, for an irradiated target area. Therefore, we conclude those fast ion peaks correspond to contamination of the surface by low-mass atoms. To avoid the inclusion of this contamination in the results, spectra and charge yields are considered only after cleaning the surface by the first nine shots. In the experiments, we average over five shots (shots no. 10-14) per target position as well as over 30 separate target positions, i.e. 150 shots in total. Shots later than shot no. 14 are excluded from our analysis to prevent target surface modification effects, which become apparent in the measurement of ion distributions after 20 shots (with a conservative safety margin). We verified that these effects do not change the depth of the hole and confirmed the linear dependence of the depth on the number of shots for the first 30 shots.

Following the charge yield experiments, the target is inspected by means of an optical microscope. The microscope has a 50x imaging objective with a numerical aperture (NA) of 0.42, yielding a depth of focus of 3 $\mu$m and enabling the determination of crater depth by straightforward optical inspection of a selected number of holes. The same microscope, equipped with a 5x imaging objective and a motorized stage for automated focus scanning to provide a complete picture of the hole, is used for an automated ablation volume determination by means of the focus variation technique\cite{danzl2011} which combines the images acquired by the microscope with computational techniques to provide 3D reconstructions of the ablated sample surfaces. A 2D Gaussian fit to the reconstructed surface profile is performed, and the integral of the fitted curve then provides an estimate of the ablated volume.

\section{\label{sec:RandD}Results \& Discussion}
\subsection*{Pulse length dependence}

\begin{figure}[b!]
\includegraphics[width=8.5cm]{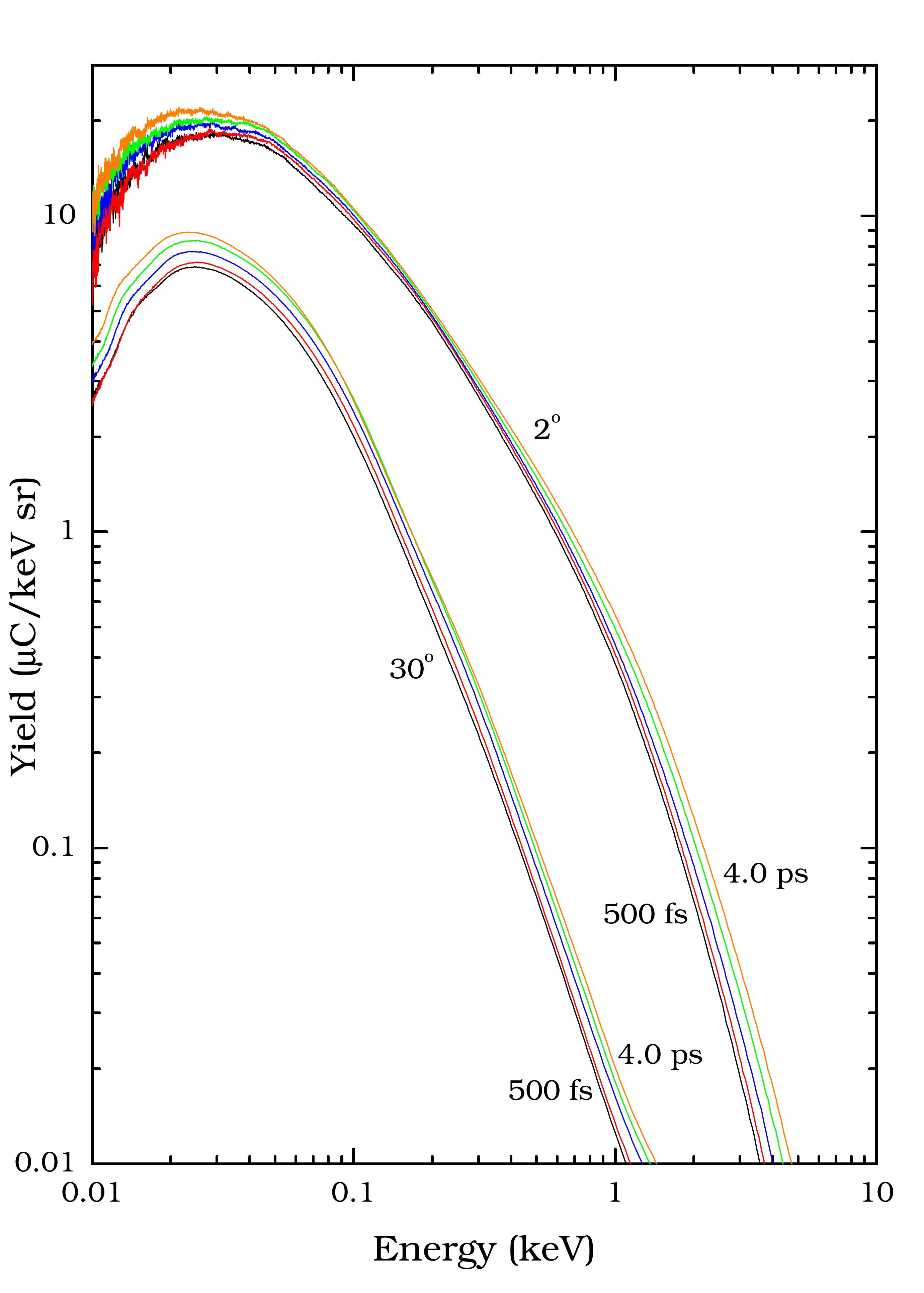}
\caption{Charge yields as a function of the ion energy for the 2\textsuperscript{$\circ$}-FC (upper set of lines) and one of the 30\textsuperscript{$\circ$}-FCs (lower set of lines). Five pulse lengths are shown: \mbox{500 fs (black)}, \mbox{1.2 ps (red)}, \mbox{2.0 ps (blue)}, \mbox{3.0 ps (green)} and \mbox{4.0 ps (orange)}. The measurements were performed with a constant peak fluence of \mbox{17 J/cm\textsuperscript{2}}.}
\label{fig:dqde_pl}
\end{figure}

\begin{figure}[]
\includegraphics[width=8.5cm]{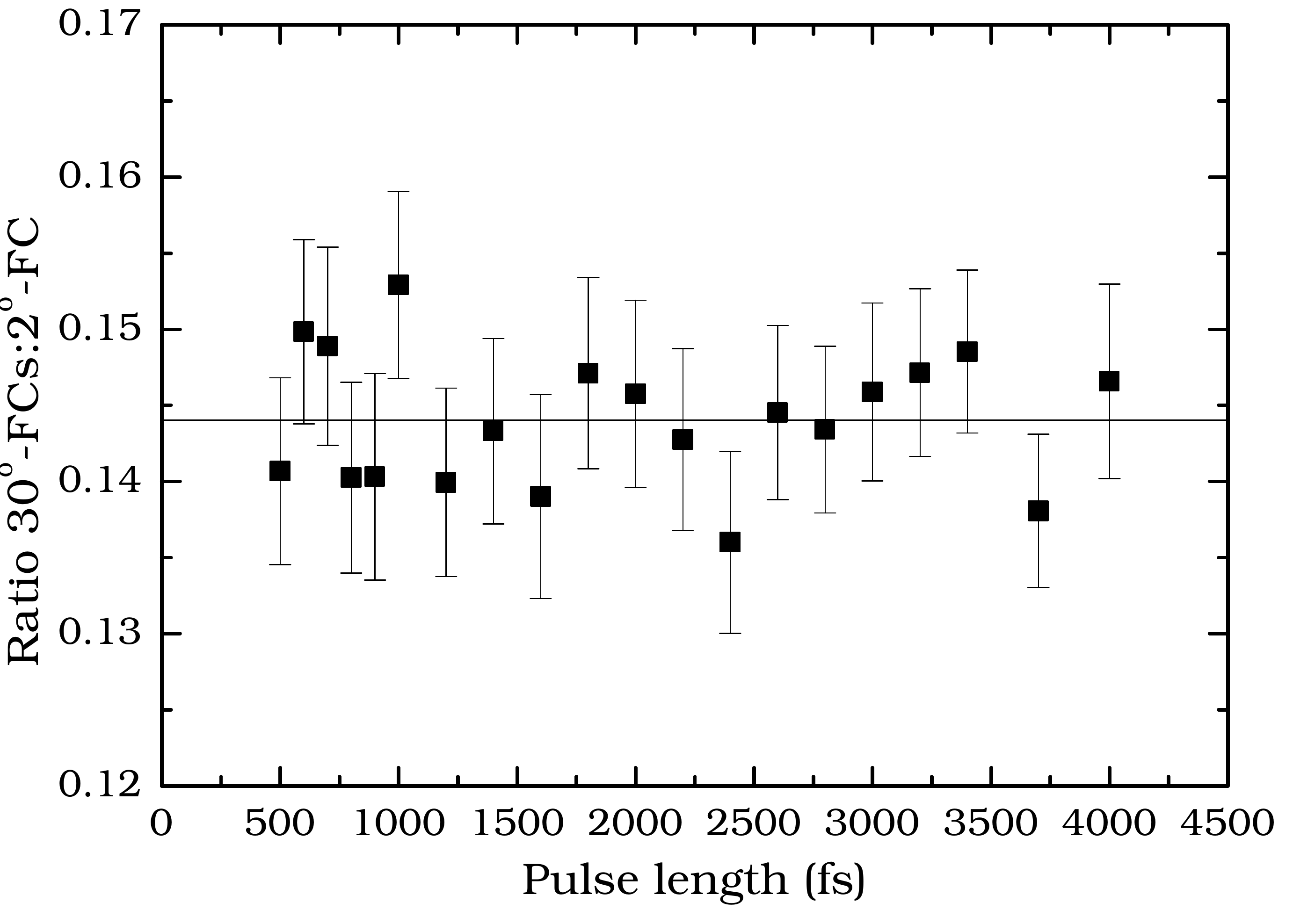}
\caption{The ratio of the yields of the \mbox{30\textsuperscript{$\circ$}-FCs} to the \mbox{2\textsuperscript{$\circ$}-FC} versus the pulse length. The black line depicts the average ratio for all pulse lengths.}
\label{fig:ratio_pl}
\end{figure}

\begin{figure}[]
\includegraphics[width=8.5cm]{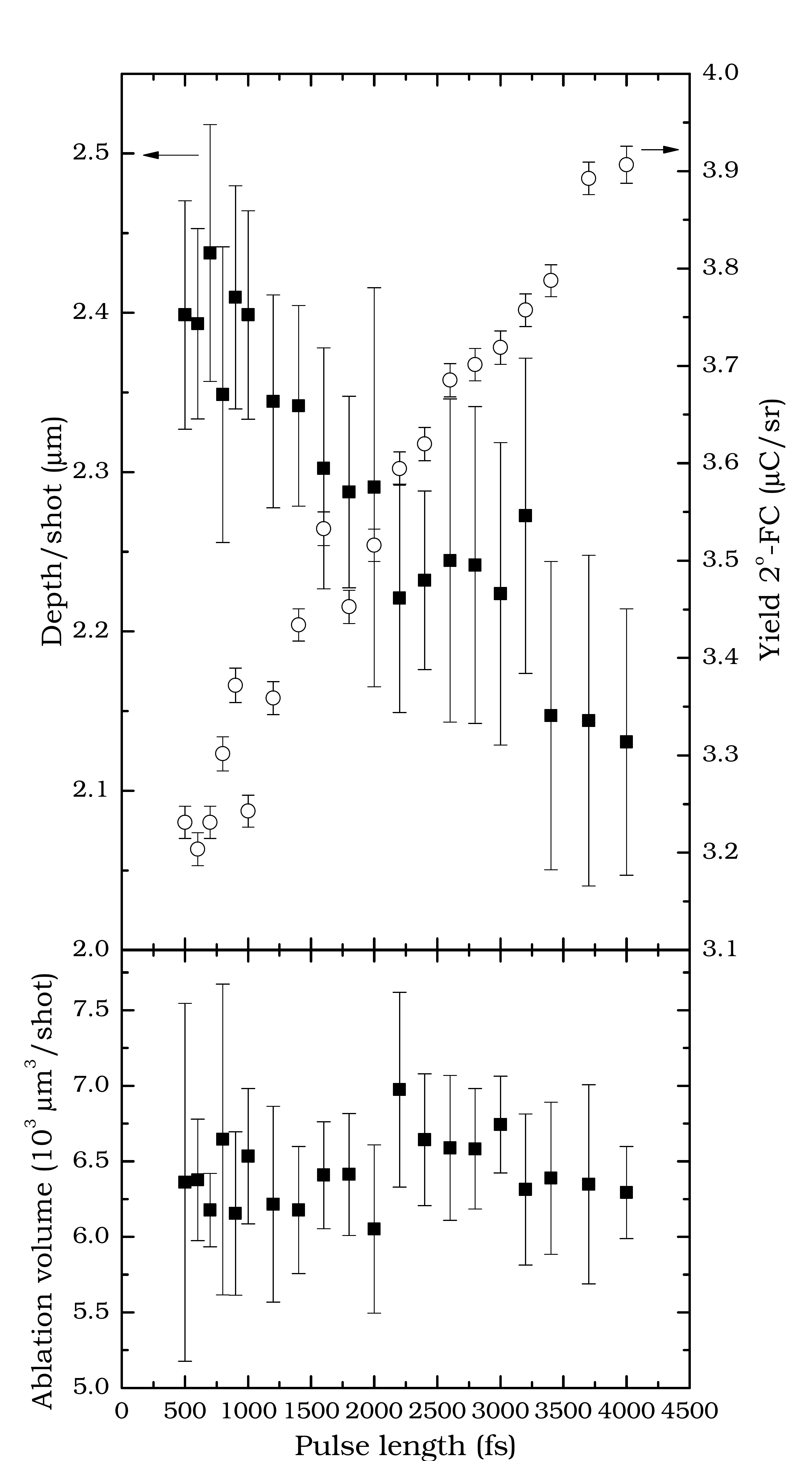}
\caption{(upper) Total charge yield at the \mbox{2\textsuperscript{$\circ$}-FC (open circles, right axis)} and the depth at the center of the holes (closed squares, left axis) as a function of pulse length. The measurements were performed at a constant peak fluence of \mbox{17 J/cm\textsuperscript{2}}. (lower) The ablation volume obtained from the focus variation technique\cite{danzl2011} as a function of pulse length at the same constant peak fluence. The error bars indicate 1-standard deviation of the mean on either side. Two data points where no reliable estimation was possible are excluded.}
\label{fig:yielddepth_pl}
\end{figure}

\begin{figure}[b]
\includegraphics[width=8.5cm]{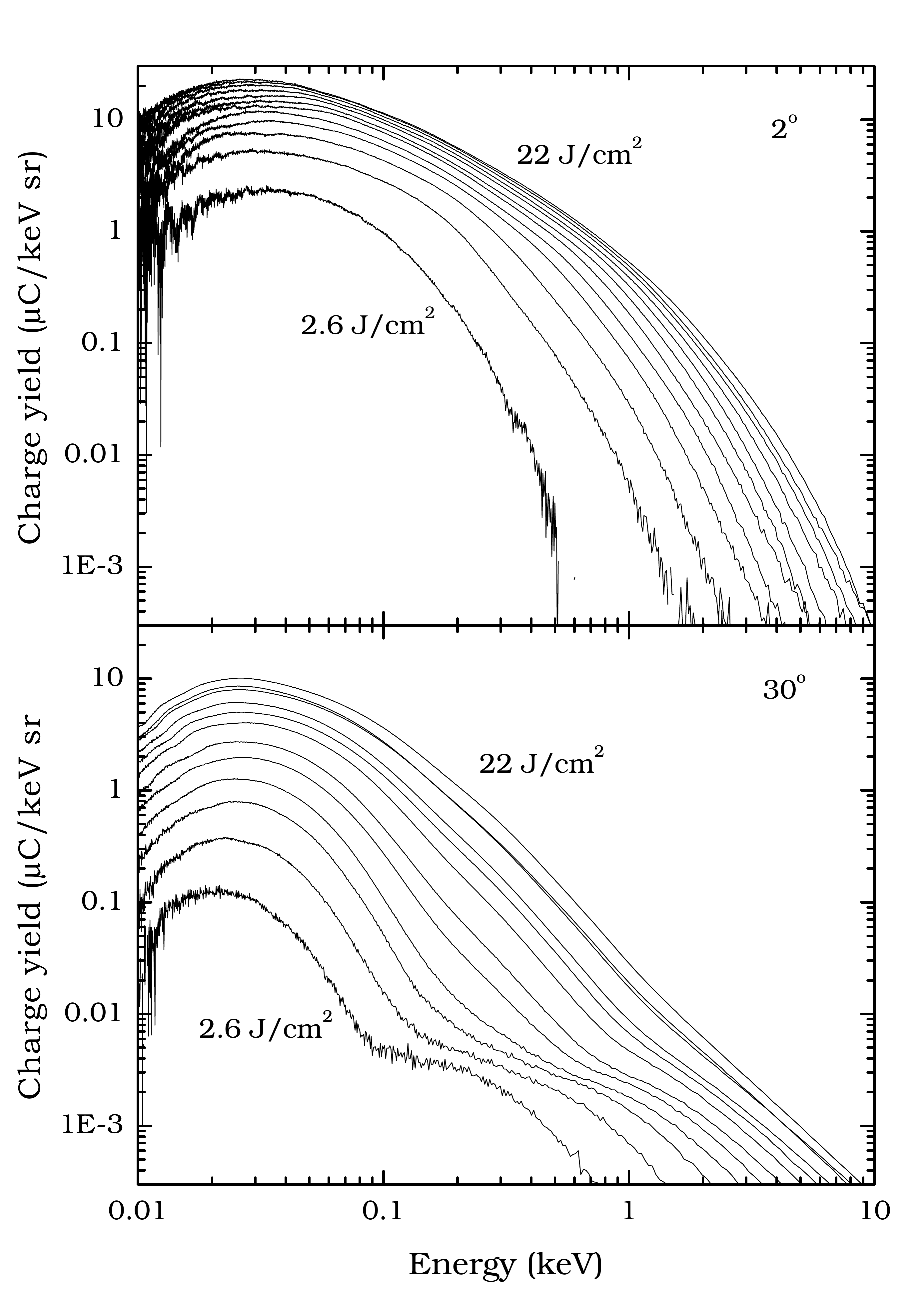}
\caption{Charge yield as a function of the ion energy for the 2\textsuperscript{$\circ$}-FC (upper panel) and one of the 30\textsuperscript{$\circ$}-FCs (lower panel) for increasing peak fluence, from 2.6 to \mbox{22 J/cm\textsuperscript{2}} in steps of \mbox{1.8 J/cm\textsuperscript{2}} at a constant pulse length of 1.0 ps.}
\label{fig:dqde_pf}
\end{figure}

\begin{figure}[]
\includegraphics[width=8.5cm]{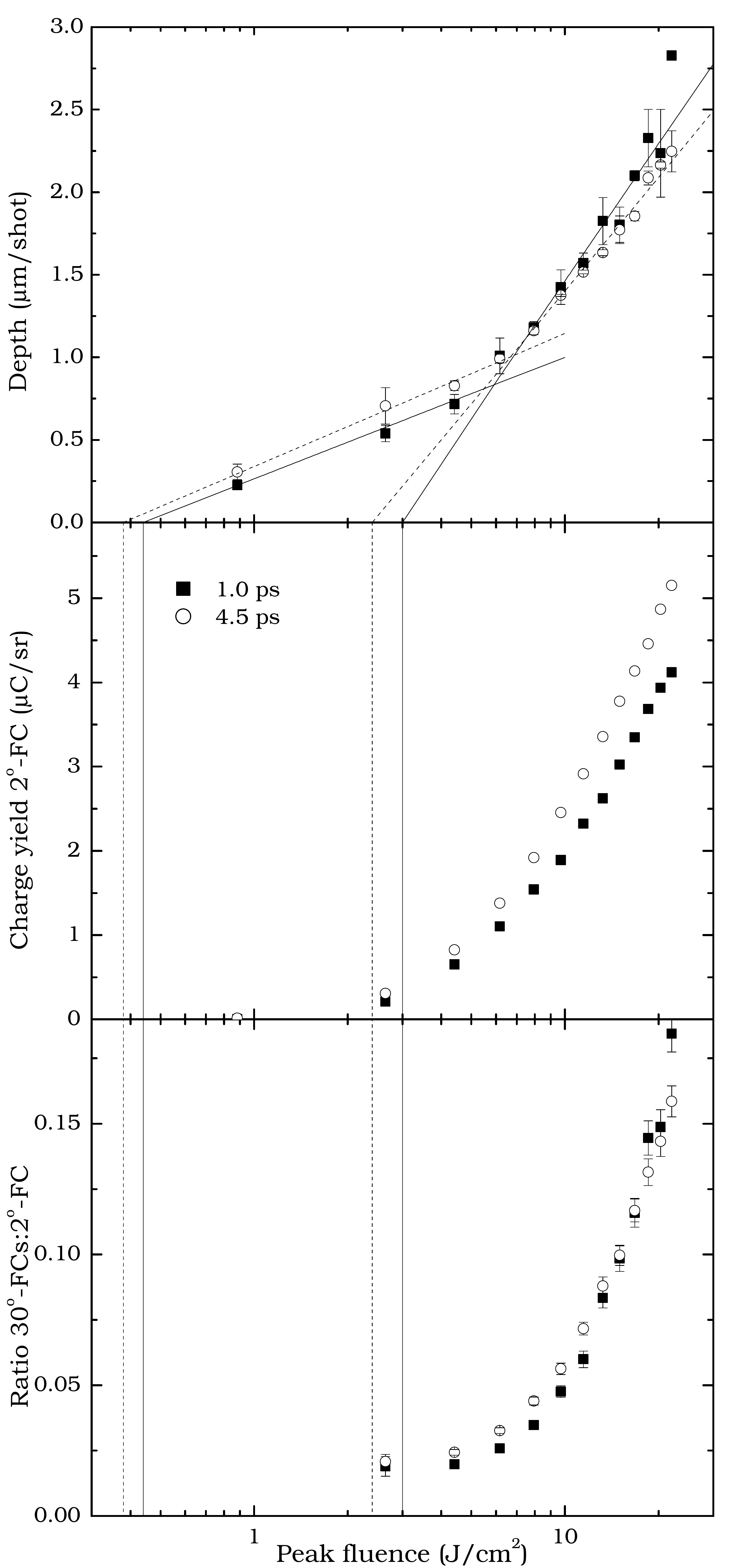}
\caption{
(upper) The ablation depth at 1.0 (filled squares) and 4.5 ps (open circles) as a function of the peak pulse fluence. The lines represent fits of equation \ref{eq:depth_pf} through the data.  Points at 6 J/cm\textsuperscript{2} are included in both fit ranges. Thresholds are \mbox{0.44 (1.0 ps)} and \mbox{0.38 J/cm\textsuperscript{2} (4.5 ps)} for the low fluence region and \mbox{3.0 (1.0 ps)} and \mbox{2.4 J/cm\textsuperscript{2} (4.5 ps)} for the high fluence region. As a reference these thresholds are also shown below (middle) Total charge yields for the \mbox{2\textsuperscript{$\circ$}-FC} at \mbox{1.0 (filled squares)} and \mbox{4.5 ps (open circles)}. The error bars are smaller than the symbol size. (lower) The ratio of the yields of the \mbox{30\textsuperscript{$\circ$}-FCs} to that of the \mbox{2\textsuperscript{$\circ$}-FC} at \mbox{1.0 (filled squares)} and \mbox{4.5 ps (open circles)}. The data point at \mbox{0.9 J/cm\textsuperscript{2}} is omitted due to low signal quality.
}
\label{fig:threepanel_pf}
\end{figure}

Fig. \ref{fig:dqde_pl} shows the charge-per-energy signal for two FCs for varying pulse lengths, ranging 500 fs to 4.0 ps. Most of the charge is due to relatively low-energy ions, in the range of 10-100 eV. The peak energy (the energy of the maximum yield) does not substantially change for changing pulse length and is located near 30 eV. Most of the ions are directed backwards with respect to laser beam, i.e. normal to the surface of the target, in line with the model of Anisimov et al. of the ion plume dynamics during laser ablation \cite{anisimov1993}. The ratio of total charge yield of the 30\textsuperscript{$\circ$}-FCs to the yield of the 2\textsuperscript{$\circ$}-FC is constant in the investigated pulse length range at a value of 0.14 (see Fig. \ref{fig:ratio_pl}), implying an angular distribution which does not depend on the pulse duration. \\

Rates of multiphoton ionization processes, in which multiple photons are directly absorbed by a single atom, are heavily dependent on the laser intensity. For laser intensities above 10\textsuperscript{14} W/cm\textsuperscript{2}, multiphoton ionization is dominant in laser ablation\cite{gamaly2011}. The maximum examined peak intensity in this work is $4.1 \times 10$\textsuperscript{13} W/cm\textsuperscript{2}, at a peak fluence of 22 J/cm\textsuperscript{2} and with a pulse length of 500 fs. Therefore, we expect that multiphoton ionization has a negligible role in the laser ablation and that the ablation and ionization in the surface is dominated by electron impact mechanisms \cite{gamaly2011}. These mechanisms are dependent on the total energy put in the system and not on the intensity, barring potential larger heat conduction losses for longer pulse lengths\cite{harzic2005,nolte1997}. The relative insensitivity of our observations to the length of the laser pulse in the studied range confirms that laser intensity itself, at a given fluence, does not play a dominant role. 

Fig. \ref{fig:dqde_pl} also shows that ion yields increase with pulse length for all ion energies. The upper panel of Fig. \ref{fig:yielddepth_pl} shows the total charge collected on the 2\textsuperscript{$\circ$}-FC together with the ablation depth for each pulse length. The charge yield increases linearly from 3.2 $\mu$C/sr at a pulse length of 500 fs to 3.9 $\mu$C/sr at 4.0 ps. In contrast, the ablation depth exhibits the opposite trend. It decreases for increasing pulse length from 2.4 (500 fs) to 2.1 (4.0 ps) $\mu$m/shot. However, the ablation volume is constant (see lower panel of Fig. \ref{fig:yielddepth_pl}), within the measurement uncertainties, because of an increase in hole radius compensates decreasing depth. The increase in accumulated charge does therefore neither have its origin in an increase of ablated material (cf. Fig. \ref{fig:yielddepth_pl}), nor in a broadening of the angular ion distribution (cf. Fig. \ref{fig:ratio_pl}). A possible explanation could be local screening of the laser light by vapor absorption\cite{harzic2005,nolte1997}. For longer pulses, more and more ablated material (ions, electrons, and neutral particles) will partially block the target surface from these laser pulses. Instead of ablating the surface, this laser light will be absorbed by the vapor. For gold, Pronko and coworkers \cite{pronko1995} used numerical simulations to show that the fraction of laser light absorbed by vapor increases from 0 to almost 20\% between 100 fs and 10 ps, respectively. This results in a decrease of the amount of ablated material because part of the laser light does not reach the target, while the vapor may be further ionized.

Concluding, we find that a longer pulse length results in a gradual increase in ionization, but a gradual decrease in the ablation depth at the center. The total amount of ablated material did not change. We observe no indications of a maximum or minimum such as found by Vinokhodov et al. \cite{vinokhodov2016,vinokhodov2016b}. This could possibly be attributed to the difference in target morphology in the comparison: Vinokhodov reported on results obtained on liquid tin droplets, whereas our work focuses on planar solid tin targets. The angular ion yield distribution is constant in the pulse length range of 500 fs to 4.0 ps. For the observed range, shortening the pulse length results in fewer ions.

\subsection*{Peak fluence dependence}

In addition to the pulse duration, experiments for a varying pulse fluence are conducted. These measurements are performed at 1.0 and \mbox{4.5 ps} pulse length. Fig. \ref{fig:dqde_pf} shows the ion spectra at 2\textsuperscript{$\circ$} and 30\textsuperscript{$\circ$} angle for all examined pulse fluences. The bulk of the ions have low energy, with a broad peak around 30 eV. More charge is collected as the pulse fluence increases for all ion energies. Particularly noticeable is the increase in the yield of high-energy ions. The yield at 40 eV ion energy increases approximately 10 times, whereas that at 400 eV increases by a factor of about 300, comparing the signals on the 2\textsuperscript{$\circ$}-FC for the highest (22 J/cm\textsuperscript{2}) and the lowest (2.6 J/cm\textsuperscript{2}) peak fluence (cf. Fig. \ref{fig:dqde_pf}). For the 30\textsuperscript{$\circ$}-FCs an additional shoulder at a higher ion energy (several hundred eV) is visible. This shoulder shifts towards higher energies for increasing pulse energy. At the high end of the fluence range the larger low-energy peak attains such heights and widths that the high-energy shoulder becomes indistinguishable from it. This high-energy feature is also visible in other ablation experiments with pulse durations in the fs-ps range \cite{amoruso2002,verhoff2012} and has been ascribed to the occurrence of an ambipolar field, resulting from a space-charge layer formed by electrons above the surface. This field accelerates some of the ions towards higher energies. It increases with temperature and the gradient of electron density \cite{doggett2011}.

Nolte and coworkers \cite{nolte1997} showed that the ablation depth has a logarithmic dependence on the laser fluence for pulse lengths up to a few ps. Typically two regions are present: a low-fluence region, in which the optical penetration of the laser light defines the ablation, and a high-fluence region, in which the electron thermal diffusion is leading. The low-fluence region has a smaller ablation depth than the high-fluence region. The precise location of the boundary between these regions is dependent on the target material and the laser characteristics.  In both regions, the depth follows the generic equation\cite{chichkov1996}

\begin{equation}
D=a\ln\bigg(\frac{F}{F_{thr}}\bigg),
\label{eq:depth_pf}
\end{equation}

\noindent in which \textit{D} is the ablation depth, \textit{a} the ablation constant, \textit{F} the laser fluence and $F_{thr}$ the threshold ablation fluence. 

We measured the depth of the hole at its center as a function of the peak fluence (see the upper panel of Fig. \ref{fig:threepanel_pf}). For both pulse lengths, the results show a clear logarithmic dependence separated in two regions, with the high-fluence region starting around 6 J/cm\textsuperscript{2}. A fit of the results for the low-fluence region shows that, within the uncertainties of the measurements, the ablation constant and threshold are the same for both pulse lengths. The ablation constant is 0.3 $\mu$m for both pulse lengths, while the ablation thresholds are 0.44 and 0.38 J/cm\textsuperscript{2} for 1.0 and 4.5 ps, respectively. In the high-fluence region the thresholds are found to be 3.0 and 2.4 J/cm\textsuperscript{2} for 1.0 and 4.5 ps, respectively. Such a decrease of the threshold is in agreement with the numerical simulations of Pronko et al. \cite{pronko1995}. The ablation constant is slightly higher for the 1.0 ps case at 1.2 $\mu$m, against the 1.0 $\mu$m found for 4.5 ps.

These ablation thresholds for tin are similar to those found with a similar experimental approach for iron by Shaheen et al. \cite{shaheen2013,shaheen2014} with 0.23 and 2.9 J/cm\textsuperscript{2} for the low- and high-fluence regions, respectively (for a lower pulse length of 130 fs). In comparison to other metals such as gold, silver, aluminum and copper, tin has higher thresholds \cite{nolte1997,furusawa1999,toftmann2013}. The high-fluence threshold of gold, for example, is reported to be 0.9 J/cm\textsuperscript{2} at roughly 150 fs\cite{furusawa1999, shaheen2014} and \mbox{1.7 J/cm\textsuperscript{2}} at almost 800 fs \cite{furusawa1999}. The theoretically expected ablation thresholds are dependent on target properties, such as optical penetration depth, thermal conductivity, and density \cite{nolte1997,furusawa1999}, and laser properties such as the pulse duration \cite{nolte1997, furusawa1999, harzic2005}. This large parameter space makes our experimental findings particularly valuable, as no straightforward predictions can be made.

The charge yield at the 2\textsuperscript{$\circ$}-FC (middle panel of Fig. \ref{fig:threepanel_pf}) increases for increasing pulse fluence, from the noise level below 0.1 to 4.1 (1.0 ps) and 5.2 $\mu$C/sr (4.5 ps). A noticeable difference with the results for the ablation depth is the higher "threshold" above which appreciable ionization is apparent in our measurements. At the lower fluences, the temperature of the surface is too low to generate an observable amount of ions and mostly neutral particles are emitted. Above a certain fluence ions are generated and the charge yield gradually increases above that fluence, following a roughly linear or logarithmic dependence. The charge yield results for both pulse lengths are very similar. In agreement with the above discussed pulse length results, the yield for the 4.5-picosecond pulses is slightly higher. As the charge yield at a certain angle is determined by several factors which are not necessarily constant for the pulse fluence, such as the volume of ablated material, angular distribution, and ionization fraction, there are no clear expectations for the fluence dependence. For these same reasons, a good comparison between studies in the available literature is also difficult to realize. Toftmann and coworkers\cite{toftmann2013} find a linear dependence for the total yield up to 2 J/cm\textsuperscript{2} whereas Amoruso et al. \cite{amoruso2000,amoruso2002} find a logarithmic dependence up to 3 J/cm\textsuperscript{2}. 

While changing the pulse length does not influence the angular ion distribution, the pulse fluence certainly does. The lower panel of Fig. \ref{fig:threepanel_pf} shows the ratio of the 30\textsuperscript{$\circ$}-FC yields to the 2\textsuperscript{$\circ$}-FC yield for both pulse lengths. The ratio increases from 0.02 near threshold to almost 0.2 at the highest fluence. At the lower fluences the ratio is fairly constant but it increases rapidly for higher fluences, indicating a rapidly broadening of the angular distribution. There is no appreciable difference between the ratios for the 1.0- and 4.5-picosecond signals. 

\begin{figure}[]
\includegraphics[width=8.5cm]{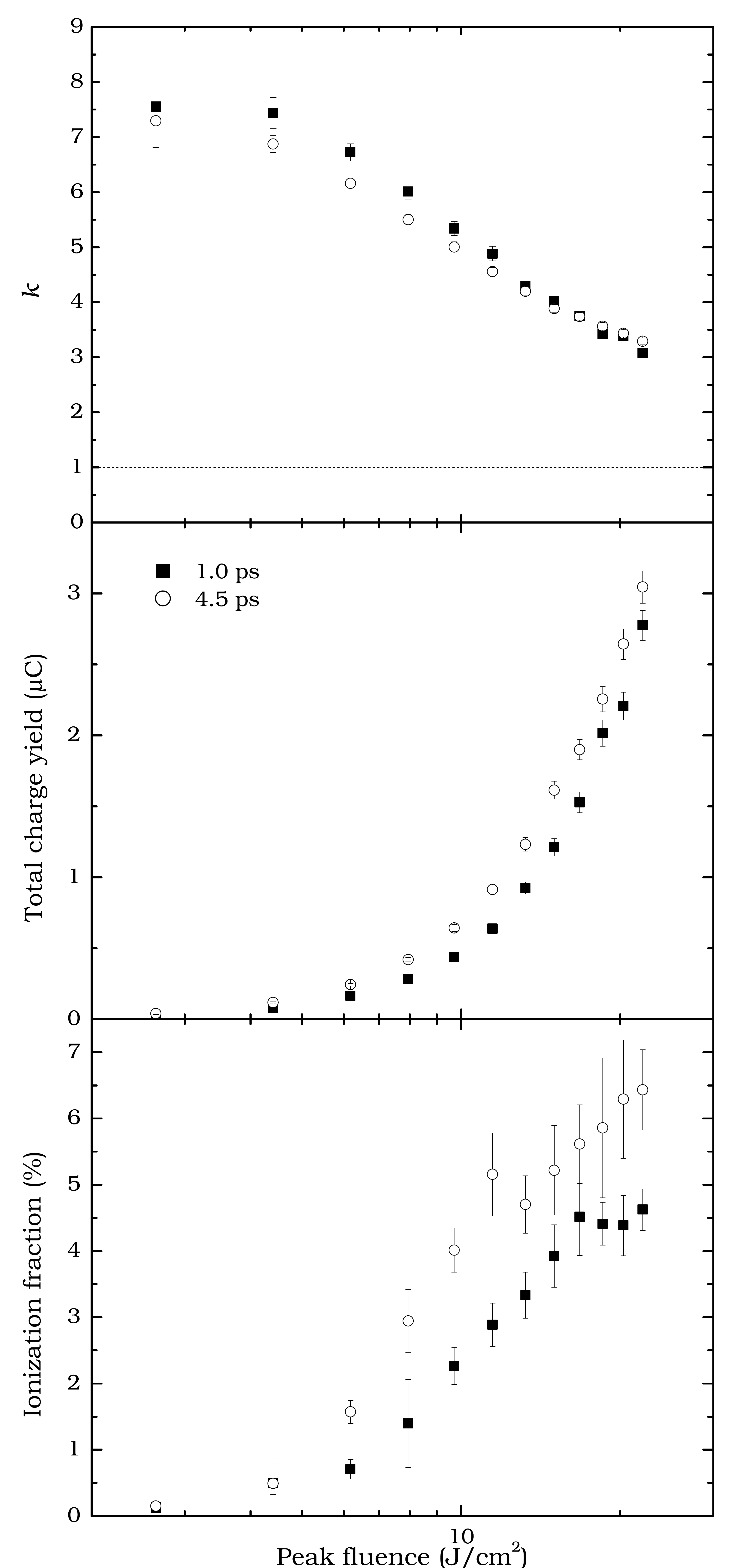}
\caption{(upper) The value of $k$ of the angular distribution (cf. equation \ref{eq:k}) versus the peak fluence for 1.0 (filled squares) and 4.5 ps (open circles) pulse length. The value of the dashed line represents the value of $k$ for which the distribution is isotropic.
(middle) The total charge yield over the whole hemisphere out of the target plane for 1.0 (filled squares) and 4.5 ps (open circles), obtained using $k$ and the total charge yield of the 2\textsuperscript{$\circ$}-FC (cf. equation \ref{eq:totalyield}).
(lower) The ionization fraction for 1.0 (filled squares) and 4.5 ps (open circles), obtained with the total charge yield and the ablation volume. The error bars indicate the 1-standard deviation of the mean, as obtained from error propagation (cf. Fig. \ref{fig:threepanel_pf}).}
\label{fig:ionfraction_pf}
\end{figure}

Following Anisimov's model\cite{anisimov1993,doggett2011}, the angular distribution of the plasma vapor from laser ablation in terms of the yield $Y(\theta)$ per unit surface at a certain polar angle $\theta$ with respect to the yield at 0\textsuperscript{$\circ$} is described by

\begin{equation}
\label{eq:k}
\frac{Y(\theta)}{Y(0)}=\bigg[\frac{1+\tan^2(\theta)}{1+k^2\tan^2(\theta)}\bigg]^{3/2},
\end{equation}

assuming cylindrical symmetry around the target normal and introducing the parameter $k$. This formula is adjusted to the hemispherical case\cite{doggett2011} from the seminal planar surface case \cite{anisimov1993}. A large value of the scaling parameter \textit{k} indicates that the angular distribution is sharply peaked in the direction along the target normal, while a $k$ equal to 1 describes a fully isotropic distribution. The values of $k$ can be obtained from the charge yield ratios depicted in Fig. \ref{fig:threepanel_pf} (lower panel) and are plotted in Fig. \ref{fig:ionfraction_pf} (upper panel). We find that $k$ decreases from roughly 8 to 3 in the examined fluence range. A similar study on the ablation of silver \cite{toftmann2013} found similarly large values for $k$ (6.2 and 4.0 depending on the axis of the elliptic spot size) at 500 fs pulse length and a fluence of 2 J/cm\textsuperscript{2}. This same study reports values for $k$ between 2 and 3 for ns-pulses, similar to studies of Thestrup et al. in the nanosecond-range \cite{thestrup2002, thestrup2003}. Those studies found a decreasing $k$ for increasing fluence, similar to our findings in the fs-ps-range. Additionally, they generally found that ion distributions from nanosecond-laser ablation are much broader than those of femtosecond-laser ablation. For tin, studies with ns-long pulses indeed found similarly broad angular ion distributions\cite{campos2010, freeman2012}.\\

To obtain the total charge yield $Y_{\textrm{total}}$ of all ions emitted from a pulse in terms of the yield at 0\textsuperscript{$\circ$}, and $k$, we integrate equation \ref{eq:k} over the relevant half hemisphere resulting in

\begin{equation}
\label{eq:totalyield}
Y_{\textrm{total}}=\frac{2\pi Y(0)}{k^{2}}.
\end{equation}

The results of the total yield are shown in Fig. \ref{fig:ionfraction_pf}. For the examined fluence range, the total yield increases from near-zero to $\sim$3 $\mu$C, corresponding to $2 \times 10^{13}$ ions, assuming singly-charged ions. The combination of increasing charge yield measured at the 2\textsuperscript{$\circ$}-FC and a broadening angular distribution results in a very rapidly increasing total charge yield. The total charge yield combined with the volume measurements enable the determination of the ionization fraction, i.e., the amount of elementary charge per atom (see lower panel of Fig. \ref{fig:ionfraction_pf}). Experiments in the fs-range, on other elements than tin, report ionization fraction values of 1\%\cite{toftmann2013} at \mbox{2 J/cm\textsuperscript{2}} (at 500 fs pulse length) to \mbox{$\sim$3-4\%\cite{ni2014}} at \mbox{5 J/cm\textsuperscript{2} (50 fs)}. We find similar values, reaching 5 and 6\% in our fluence range for 1.0 and 4.5 ps respectively. This is significantly lower than the ionization fraction of several 10\% observed in nanosecond laser ablation (at fluences of \mbox{$\sim$ 2 J/cm\textsuperscript{2})\cite{thestrup2002, doggett2011}}.

\section{\label{sec:conclusions}Conclusions}

We have studied the influence of two laser parameters on the ion charge yield and energy distribution, as well as the ablation depth and volume. A high-energy ion peak is visible for low fluences, in agreement with the available literature. Variation of the pulse duration from 500 to 4000 fs results in a small increase of the ion charge yield, while the ablation depth decreases slightly. A possible explanation is the screening of the target by the plasma plume. The total ablation volume remains constant. Interestingly, we do not observe the abrupt changes in either depth or ion yield that were hinted at in refs. \onlinecite{vinokhodov2016,vinokhodov2016b}. The ion yield angular distribution does not change appreciably as a function of pulse length. The ablation depth follows a two-region logarithmic dependence on laser pulse peak fluence, in agreement with the existing theory. We find ablation thresholds of 0.44 (at a pulse length of 1.0 ps) and \mbox{0.38 J/cm\textsuperscript{2} (4.5 ps)} for the low-fluence region and \mbox{3.0 (1.0 ps)} and \mbox{2.4 J/cm\textsuperscript{2} (4.5 ps)} for the high-fluence region, close to literature values for other metallic elements. The "threshold" at which ionization is apparent is higher, from there on the ion charge yield increases in step with fluence. The angular distribution is sharply peaked backwards along the target normal at the lower fluences, but rapidly broadens for the higher fluences. The total ionization fraction increases gradually and monotonically with the fluence to a maximum of 5-6\%, which is substantially lower than typical values for nanosecond-laser ablation. These results are of particular interest for the possible utilization of fs-ps laser systems in plasma sources of EUV light for next-generation nanolithography.

\section*{Acknowledgements}
We want to thank the Ultrafast Spectroscopy group of professor Huib Bakker at AMOLF Amsterdam for the opportunity to use the IRIS laser system and their laboratory. Furthermore we would like to thank the AMOLF and ARCNL workshops and technicians for the aid provided during the experiments. S.W. acknowledges funding from the European Research Council (ERC-StG 637476).

\end{document}